# The Era of Extremely Large Optical Telescopes I: The ELT

Priya Hasan*

May 21, 2026

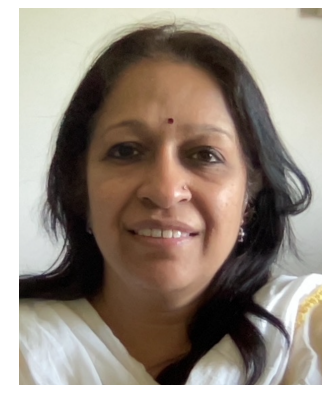

Dr. Priya Hasan is an Assistant Professor of Physics at Maulana Azad National Urdu University, Hyderabad. Her areas of interest are observational astronomy, star formation, star clusters, and galaxies. Beyond her research, Dr. Hasan is deeply committed to astronomy education and public outreach. She serves as the Regional Astronomy Education Coordinator for the IAU's Office of Astronomy Education (OAE) in India and is a former co-Chair of the IAU's Women in Astronomy Working Group.

**Abstract**

The advent of Extremely Large Telescopes (ELTs)—ground-based optical/infrared observatories with primary mirrors exceeding 20 m—heralds a transformative epoch in observational astronomy. This article examines the dawn of this new era, and the three upcoming facilities in the optical/infrared bands: the Giant Magellan Telescope (GMT), the Thirty Meter Telescope (TMT), and the European Extremely Large Telescope (ELT). This article will focus on the ELT, while a sequel will cover GMT and TMT. We describe the key technological breakthroughs enabling its construction, most notably segmented mirror design, advanced adaptive optics (AO), and laser guide star systems. These innovations will deliver more than an order of magnitude leap in light-gathering area and spatial resolution, providing image sharpness exceeding that of space-based telescopes for wide-field observations.

The scientific impact of the ELT is profound and multifaceted. We discuss its inception and construction milestones and explore their potential to directly image and characterize the atmospheres of Earth-like exoplanets, searching for biosignatures, trace the formation of the first stars, galaxies, and supermassive black holes. Furthermore, ELT will serve as an

*Orcid Id 0000-0002-8156-nat6940

unprecedented probe of fundamental physics, offering stringent new tests for dark energy and dark matter models by measuring the universe's expansion history and the chemical evolution of its earliest structures. This paper concludes that ELT are not mere incremental improvements but foundational instruments that will redefine the frontiers of astrophysics, address some of science's most enduring questions, and inevitably lead to discoveries beyond current prediction.

*Where there is an observatory and a telescope, we expect that any eyes will see new worlds at once.*
*Henry David Thoreau*

# 1 A Brief History of the Telescope Apertures

The quest for a bigger aperture—the diameter of a telescope's primary light-gathering lens or mirror—is the holy grail of observational astronomy. More aperture means more light, which means fainter objects, finer detail, and a deeper view into the past[1] and details of the cosmos[2]. **A Brief History of Telescopes**

The Refractor Era (1609 – c. 1750)
The Great Refractor & Early Reflector Era (c. 1750 – 1910)
The Modern Glass-Mirror Reflector Era (c. 1900 – 1990)
The Segmented Mirror  Adaptive Optics Era (c. 1990 – 2010)
The Extremely Large Telescope (ELT) Era (2010s – Future)

The story begins not with a quest for size, but for simple magnification. In 1609, Galileo Galilei used a simple refracting telescope with an aperture of about 37 mm (1.5 inches). Despite its size, it revolutionized our cosmic view, discovering Jupiter's moons, lunar craters, and the Milky Way's stellar nature. For a century, these 'spyglasses' remained small due to poor lens quality and chromatic aberration (color fringing). Christiaan Huygens built long, tubeless aerial telescopes to minimize aberrations, with lenses up to 120 mm (5 inches). These unwieldy instruments discovered Saturn's moon Titan.

The 18th century saw the mastery of lens-making, leading to the 'Great Refractors'. Tired of refractor limitations, William Herschel built large reflecting telescopes building his 'Great 40-foot' telescope with a 1.2 m (48-inch) speculum metal mirror. It was monumental but clumsy.

Herschel's real workhorse was a 47 cm (18.7-inch) reflector, which he used to discover Uranus and map thousands of nebulae.**Key disadvantages of refractors**:

[1] The finite speed of light is what lets us look back in time. As the Sun is 8 light minutes away, light we receive left the Sun 8 minutes ago and lets us see the Sun as it was 8 minuts ago. Similarly, if we look at very distant objects that are millions of light years away, we see the object as it was millions of years ago. Hence we can see the universe in its early stages.

[2] An increase in telescope diameter leads to a smaller Field of View (FOV). This is given by: Real FOV = Eyepiece FOV/Magnification. So to observe the full moon ($0.5^0$) for a telescope with eyepiece with a FOV of 60', the magnification will need to be 120X

Figure 1: Aperture diameters as a function of commissioning dates for major telescopes. Open circles are refractors and filled circles are reflectors; small filled circles are 20th century telescopes with ; D ≥ 1 crosses, instruments with shortcomings; diamonds, future instruments, including five ELTs. Image credit: [1]

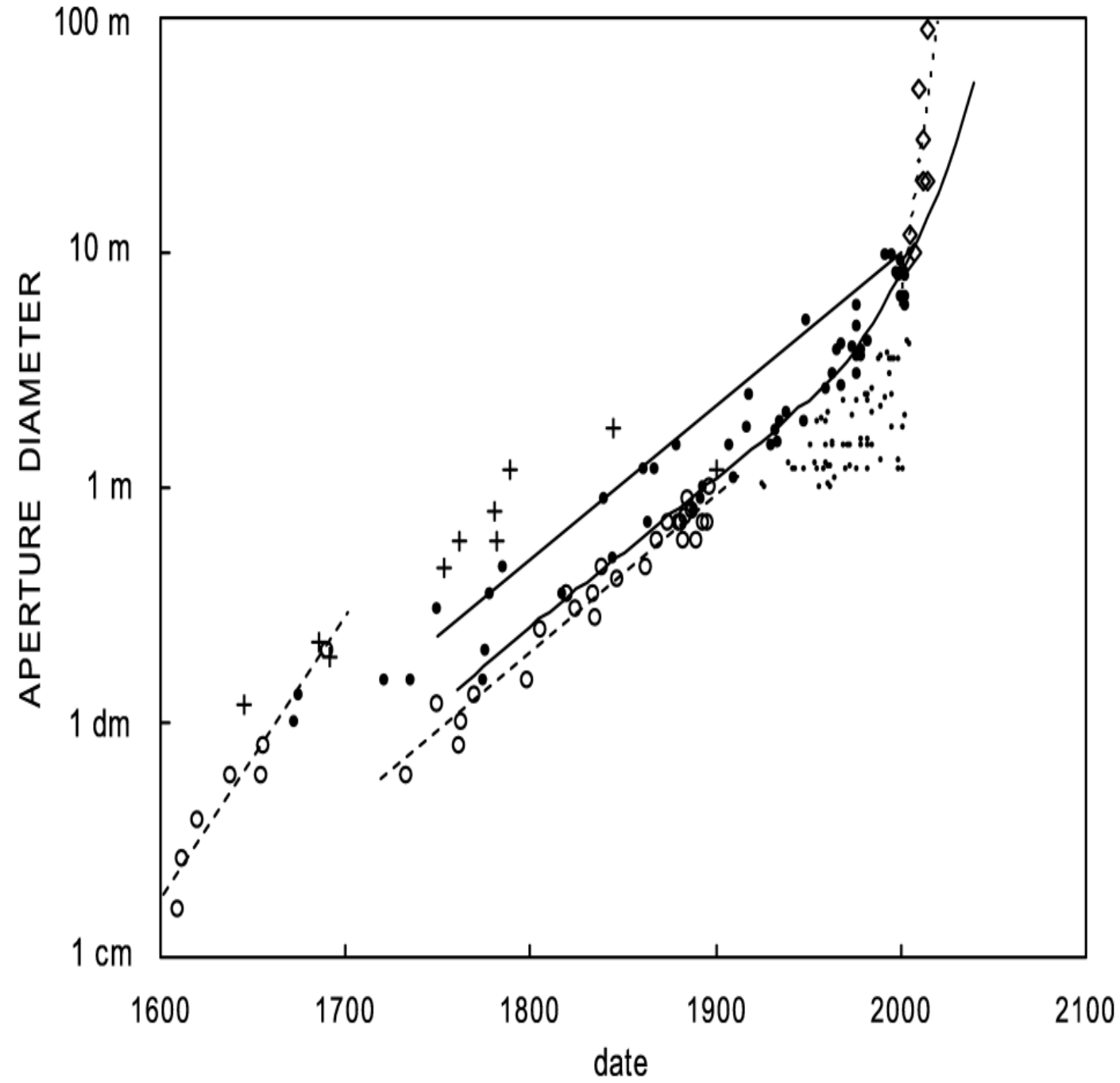

**Chromatic Aberration:** Colors do not focus at the same point, creating color fringes

**Size/Weight Limitations:** Large lenses are difficult to manufacture, expensive, heavy

**Physical Limitations:** Very long tubes required for high magnification, cumbersome and requiring heavy, expensive mounts.

**Light Loss:** Absorbtion by glass

**Spherical Aberration:** Distortions due to lens imperfections. The late 19th century saw a battle of giant lenses, culminating in the 1.02 m (40-inch) refractor at Yerkes Observatory (1897), which remains the largest successful refractor ever built.

The limits of glass lenses (sagging under their own weight, chromatic aberration) had been reached.

The switch from metal to glass mirrors, coated with reflective aluminum, changed everything. Glass mirrors could be larger, lighter, and better sup-ported. In 1917, at Mount Wilson Observatory, the 2.54 m (100-inch) Hooker Telescope reflector was the first to definitively prove that 'spiral nebulae' were distant galaxies beyond our Milky Way (Edwin Hubble, 1924) and most importantly, redefined our cosmic scale with Hubble's law and the expansion of the universe. The Hale Telescope (1948), was the 5.08 m (200-inch) giant at Palomar Observatory reigned as the world's largest effective telescope for 45 years. It was a marvel of engineering and pushed the limits of what a single, rigid mirror could be. In astronomy, seeing describes how much Earth's atmosphere distorts telescopic images, affecting their clarity and sharpness. It is governed by turbulence and temperature fluctuations in the air: poor seeing causes stars to blur and twinkle, while good seeing yields steady, high-resolution views. Seeing is typically quantified in arcseconds, which is the diameter of a star image. Excellent seeing is $\leq$ 0.4", good is 0.4" - 0.9" , and average is 1.0"-2.0".It mapped the universe to unprecedented depths and discovered quasars. The Soviet BTA-6 (Large Altazimuth Telescope) is a -m (20 ft) aperture optical telescope at the Special Astrophysical Observatory, Russia. The BTA-6 achieved first light in late 1975, making it the largest telescope in the world until 1990, when it was surpassed by the partially constructed Keck 1. It pioneered the technique, now standard in large astronomical telescopes, of using an altazimuth mount with a computer-controlled derotator.

To grow beyond 8 meters, engineers developed new concepts. The two Keck Telescopes built in 1993 and 1996 were game-changers. Each Keck telescope uses a 10 m primary mirror composed of 36 hexagonal segments, actively controlled by computers to act as a single surface. This segmented mirror design broke the single-mirror barrier and set the template for all future giants. The Very Large Telescope (VLT, 1998-2000) was made of four independent 8.2 m unit telescopes that can also combine light as an interferometer, achieving the resolution of a 130 m telescope. At 8.4 m, the largest monolithic primary mirrors in operation today are the Large Binocular Telescope (LBT) in Arizona and the upcoming Vera Rubin Observatory in Chile. For larger mirrors, segmented mirrors are the way ahead. The maximum resolution a telescope can provide due to the diffraction caused by the wave nature of light is $\frac{1.22\lambda}{D}$. This is often called the 'diffraction limit' to seeing. To add on to this is the effect of the atmosphere and turbulance that further distorts images and leads to bad seeing. Space telescopes have an advantage of no atmosphere, leading to 'diffraction-limited' seeing. Ground-based telescopes improve their seeing by a parallel revolution called Adaptive and Active Optics (AO). This technology uses deformable mirrors to correct, in real-time, for the blurring caused by Earth's atmosphere, allowing ground-based telescopes to often outperform Hubble and now James Webb in sharpness. Table 1 describes the basic differences between Active and Adaptive Optics.

The largest single (monolithic) primary mirrors currently in operation are 8.4 m in diameter, utilized by the Large Binocular Telescope (LBT) in Arizona and the upcoming Vera C. Rubin Observatory in Chile. The Subaru Telescope's 8.2 m mirror held the single largest mirror title before LBT and remains one of the largest. There is a technological limit for primary mirrors made of a

Table 1: Differences between Active and Adaptive Optics

| Feature | Active Optics | Adaptive Optics |
|---|---|---|
| Correction Target | Telescope shape (gravity, thermal) | Atmospheric turbulence |
| Speed | Slow ($\approx$1 Hz or less) | Fast (100–1000 Hz) |
| Component | Actuators on primary/secondary | Deformable mirror |
| Main Use | Maintaining mirror figure | High-res imaging/AO |

single rigid piece of glass. Such non-segmented, or monolithic mirrors have not be constructed larger than 8.4 m in diameter. **Monolithic vs. Segmented Telescope Mirrors**

**Monolithic (Single-Piece) Mirrors**

Largest in operation are 8.4 m (Large Binocular Telescope (LBT) and the upcoming Vera Rubin Observatory) Also, Subaru 8.2 m

**Segmented Mirrors (For Larger Apertures)**

Segmented mirrors are composed of many smaller, usually hexagonal pieces working together as one large surface.

Active Alignment: A computer-controlled active optics system uses actuators built into the support cell to constantly align the segments, allowing them to function as a single, unified mirror.

For mirrors larger than 8 m, the mirrors are segmented, that is, they are composed of many smaller, hexagonal pieces working together as one large surface. This design allows for telescopes with much larger total apertures. However, segmented mirrors do have the drawback that each segment may require some precise asymmetrical shape, and rely on a complicated computer-controlled mounting system. All of the segments also cause diffraction effects in the final image.

Some examples of telescopes and their number of segments are : Keck I & II (36), Gran Telescopio Canarias (GTC) (36), Hobby Eberly Telescope (HET) (91), Southern African Large Telescope (SALT) (91), Large Sky Area Multi-Object Fibre Spectroscopic Telescope (LAMOST) (24) and the James Webb Space Telescope (JWST) (18). In these telescopes, the mirror segments must be polished to a precise shape. A computer-controlled active optics system, utilizing actuators in the support cell, then constantly aligns them to function as a single, unified mirror.

The future now is the Era of Extremely Large Telescopes (ELT) (2020s – Present & Future) hook. The three primary proposed/under-construction ground-based Extremely Large Telescopes (ELTs) are the Extremely Large Telescope (ELT) in Chile (39.5m), the Thirty Meter Telescope (TMT) in Hawaii (30 m), and the Giant Magellan Telescope (GMT) in Chile (effective 25.4m). **Advancements in image quality and telescope technology:**

**Overcoming Atmospheric Interference**

Space Telescopes: To avoid atmospheric distortion, place telescopes in space, for example, The James Webb Space Telescope (JWST)

**Revolution in Resolution: Interferometry**

**Very Large Array (VLA):** Radio interferometry using 27 moveable antennas in New Mexico.

**Event Horizon Telescope (EHT):** An international collaboration using Very Long Baseline Interferometry (VLBI) that imaged the supermassive black hole at the center of galaxy M87*.

**Atacama Large Millimeter/submillimeter Array (ALMA):** A high-altitude array in Chile consisting of 66 antennas **Square Kilometer Array (SKA):** A next-generation massive radio telescope array currently under construction.

The European Extremely Large Telescope (ELT) is the current record-holder in development[3]. A 39 m segmented mirror (798 segments), under construction in Chile. First light expected in 2028. The GMT uses seven of the world's largest monolithic mirrors (8.4 m each) to form an effective 25.4 m aperture. First light expected in 2030. The Thirty Meter Telescope (TMT) is a 30 m segmented mirror telescope (492 segments). There were some other ELTs that have been shelved: the Very Large Optical Telescope (VLOT) by Canadian (20 m) and the Overwhelmingly Large Telescope (OWL) at 100 m.

The parallel revolution in improving image quality by eliminating the effect of the atmosphere are space telescopes. The James Webb Space Telescope's 6.5 m segmented mirror is the largest in space, limited by launch vehicle fairings. Interferometry in Radio telescopes was achieved at the Very Large Array (VLA), which is a 27-antenna array in New Mexico that provides 351 baselines. The Event Horizon Telescope (EHT) is an international network of telescopes that used VLBI (Very Long Baseline Interferometry) to image the M87* black hole. The Atacama Large Millimeter/submillimeter Array (ALMA) is a high-altitude array for high-frequency radio observations. The Square Kilometer Array (SKA) is a next-generation, massive radio telescope array, promising unprecedented sensitivity. Liquid mirror telescopes, telescopes on the Moon and 'solar gravitational lens' telescope that uses the Sun's gravity as a lens could, in theory, create apertures measured in kilometers or even AU.

From Galileo's 37 mm spyglass to the 39 m behemoths under construction, the light-gathering area of our greatest telescopes has increased by a factor of over one million. This relentless drive for more photons has been the engine of cosmic discovery, repeatedly redefining our place in the universe and promising answers to questions we are only now learning to ask. The history of the aperture is the history of astronomy itself.

This article will focus an ELT, and an accompanying paper by the author will discuss GMT and TMT.

[3]Earlier proposed as the Overwhelmingly Large Telescope (OWL) of 100m.

Figure 2: The Extremely Large Telescope under construction in Chile. (Image credit:

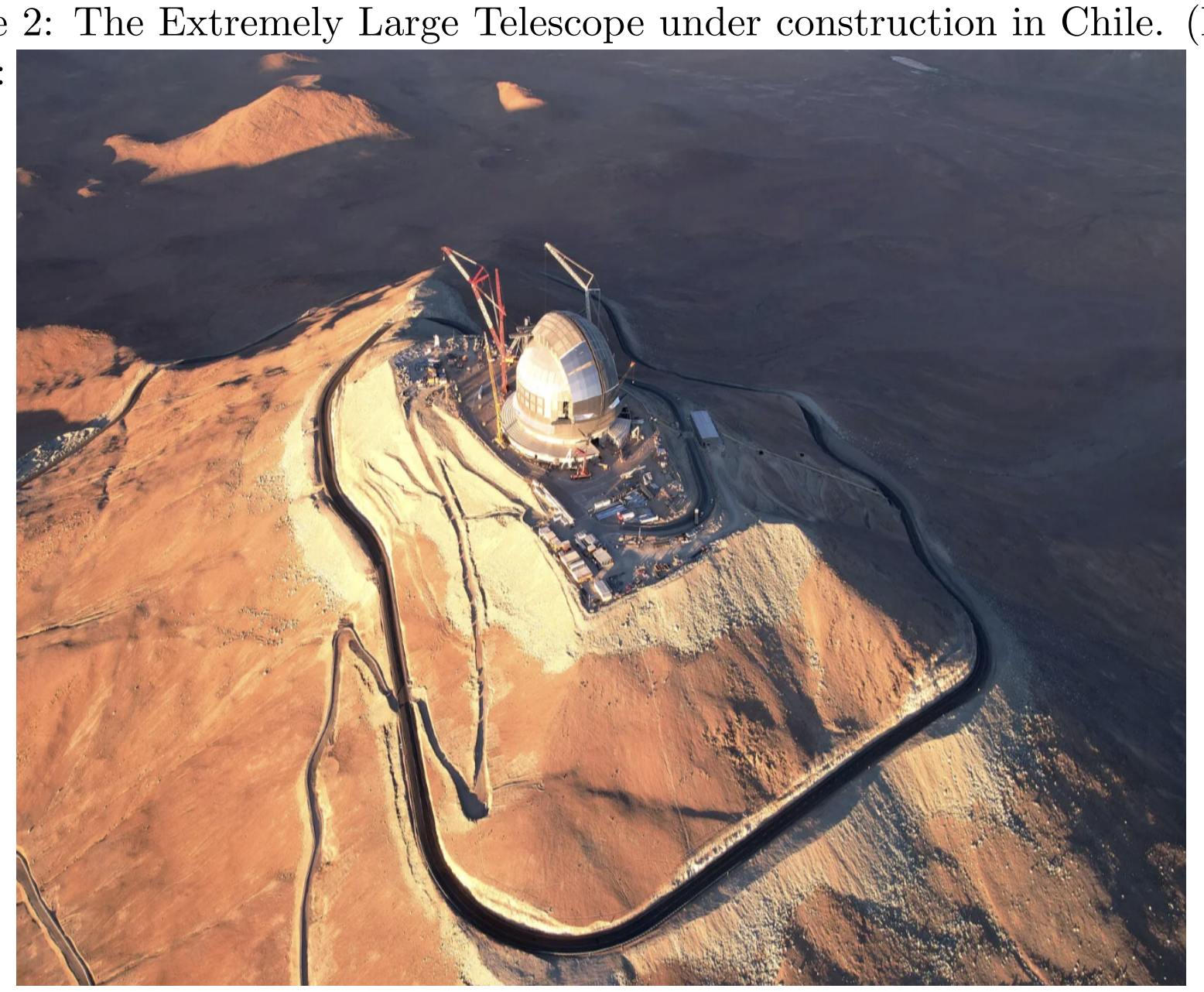

## 2 The European Extremely Large Telescope

**Future and Theoretical Concepts**

**Liquid Mirror:** Telescopes using rotating reflective liquid (such as mercury) as a primary mirror.

**Lunar Telescopes:** Telescopes on the Moon

**Solar Gravitational Lens:** Using the Sun's gravity as a natural lens to focus light enabling direct imaging of exoplanet surfaces.

Every great telescope begins with the right location. The ELT's home is Cerro Armazones, a 10,000-foot peak in Chile's Atacama Desert, selected for its near-perfect conditions: 300 crystal-clear nights per year, a commanding view of the entire southern sky and a large portion of the northern hemisphere and exceptionally low wind, dust, and temperature fluctuations [4].

Construction work on the ELT site started in June 2014. To prepare the site, engineers executed a monumental feat of earthworks, blasting away 200,000 cubic meters of rock to create a vast, flat platform anchored to the mountain's bedrock. The foundation incorporates 118 seismic isolators —sophisticated dampers that will keep the telescope perfectly still during the region's frequent earthquakes (see Fig 2).

Protecting the ELT's delicate optics is a colossal dome wider than a professional soccer field. **Some ELT Highlights: Dome Rotation:** The 5,800-ton upper dome rotates smoothly on 36 giant trolleys, tracking the telescope at speeds up to $2^0/sec$ to switch between targets in under five minutes.

[4]https://elt.eso.org/

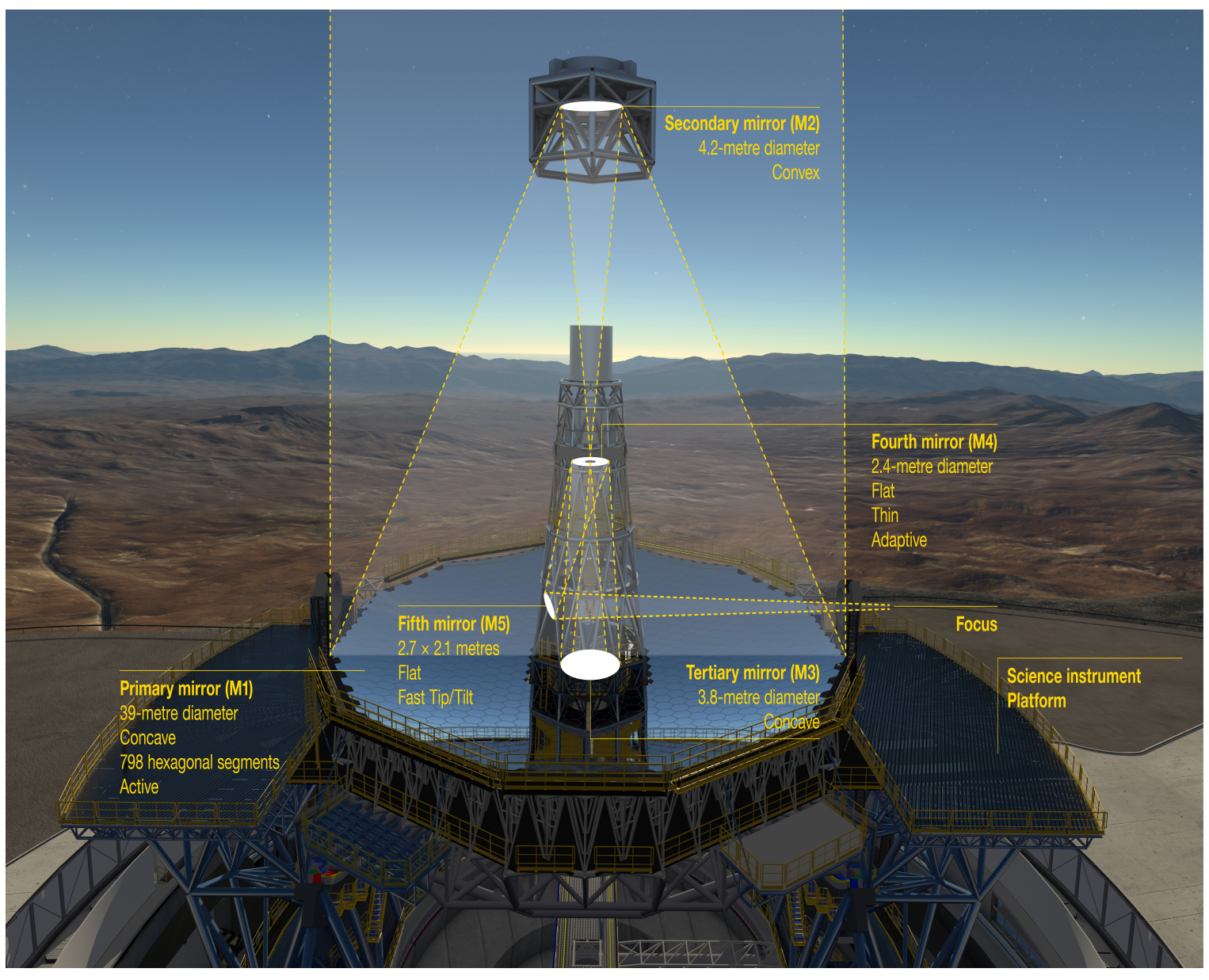

**Dome Aperture:** At night, two 55 m tall doors open with no glass barrier to prevent distorting temperature layers.

**Climate Control:** During the day, multi-layer insulation and a massive HVAC system

**Telescope Structure:** A 4,600-ton altitude-azimuth mount, rotates on a near-frictionless film of hydrostatic oil bearings .This is no passive shell, but an active environmental control system. The 5,800-ton upper dome rotates smoothly on 36 giant trolleys, tracking the telescope at speeds up to 2 degrees per second. This allows the observatory to switch between celestial targets in under five minutes.

At night, two 55-meter-tall doors, each weighing as much as a commercial jet, swing open. There is no glass barrier, as it would create distorting temperature layers. The telescope observes the sky directly. During the day, multi-layer insulation and a massive Heating, Ventilation, and Air Conditioning systems (HVAC) system actively cool the interior, ensuring the entire structure matches the outside air temperature by nightfall to prevent heat distortions that would ruin observations. The telescope passed the halfway point in its development and construction in July 2023, with the expected completion and first light set for March 2029.

Inside the dome rests the 4,600-ton telescope structure. This altitude-azimuth mount rotates with exquisite smoothness on a near-frictionless film of hydrostatic oil bearings. This engineering marvel must be both incredibly rigid and exquisitely precise.

The ELT's true brilliance is its five-mirror optical system, a cascade of engineering that refines starlight into a perfect scientific image. More details can be found in [3, 4]. **ELT Mirror System:**

**Primary Mirror (M1):** A 39-meter segmented giant made of 798 hexagonal segments, 1.4 m wide, polished to atomic-scale smoothness and silver-coated. An active support system with thousands of actuators adjusts each segment thousands of times per second to maintain a flawless surface.

**Secondary (M2) & Tertiary (M3) Mirrors:** M2, at 4.2 m , is the largest secondary mirror ever built.

**Adaptive Mirror (M4):** The world's largest adaptive optics mirror, composed of six flexible glass shells just 2mm thick. It is deformed 1,000 times per second by over 5,000 actuators to cancel out atmospheric turbulence in real time.

**Tip-Tilt Mirror (M5):** Makes rapid, fine adjustments to counteract residual vibrations

**Adaptive Optics System:** M4 & M5 with AO enables images up to 16 times sharper .

- The Primary Mirror (M1): The 39-meter Segmented giantmirror is a mosaic of 798 hexagonal segments, each 1.4 meters wide.

  Each segment is polished to atomic-scale smoothness (tens of nanometers) and coated in a silver layer just a few atoms thick. An active support system with thousands of actuators constantly adjusts each segment thousands of times per second, making the entire array behave as a single, flawless surface.

- The Secondary (M2) & Tertiary (M3) Mirrors: M2, at 4.2 meters, 3.5 tons, is the largest secondary mirror ever built, with a complex hyperbolic shape polished to nanometer perfection. Together with M1 and M3, they form an 'anastigmat' system that mathematically cancels out key optical distortions across a wide field of view.

- The Adaptive Mirror (M4): The Real-Time Corrector is the world's largest adaptive optics mirror. Composed of six flexible glass shells just 2mm thick, it is deformed 1,000 times per second by over 5,000 actuators. It actively warps to cancel out atmospheric turbulence in real time, turning a blurry twinkle into a sharp point of light.

- The Tip-Tilt Mirror (M5): The Ultimate Stabilizer M5 makes rapid, fine adjustments to counteract any residual vibration from wind or the telescope itself, locking the image in place for razor-sharp, long exposures.

Together, M4 and M5 form an adaptive optics system built directly into the telescope's core optics, enabling it to produce images up to 16 times sharper than conventional ground-based telescopes.

The corrected light is routed to a suite of first-generation instruments, each a masterpiece designed for specific cosmic quests: **The first-generation ELT** **instruments are : MICADO** delivers ultra-sharp near-infrared imaging to study faint objects and exoplanets; **HARMONI** analyzes the light of stars and galaxies to reveal their composition and motion; **METIS** probes the dusty birthplaces of stars and planets in the mid-infrared; **MOSAIC** observes hundreds of galaxies at once to map the Universe in 3D; and **ANDES** performs ultra-precise spectroscopy to search for biosignatures in Earth-like exoplanet atmospheres.

- MICADO: A powerful near-infrared camera to capture ultra-sharp images and hunt for faint exoplanets.
- HARMONI: A spectrograph to dissect the light of stars and galaxies, revealing their composition and motion.
- METIS: A mid-infrared camera/spectrograph to peer through dust and study the birth of stars and planets.
- MOSAIC: A multi-object spectrograph capable of analyzing hundreds of galaxies simultaneously, mapping the universe in 3D.
- ANDES: An ultra-precise spectrograph designed to search for chemical biosignatures in the atmospheres of Earth-like exoplanets.

To guide its adaptive optics, the ELT cannot always rely on natural stars. Its solution? Create its own.

The telescope will fire up to eight powerful orange lasers into the sky, exciting a layer of sodium atoms 90 km high to create artificial laser guide stars. Multiple lasers allow for tomographic adaptive optics, building a 3D map of atmospheric turbulence.

Processing this torrent of data requires a supercomputer that performs a monumental task: it takes 10,000 measurements, solves complex algorithms, and commands thousands of actuator adjustments—all up to 1,000 times every second.

The ELT is progressing on a disciplined schedule, with mirror installation set for 2027 and first light targeted for 2028. With an estimated cost of $ 1.5 billion, it stands as a model of ambitious yet focused engineering.

Upon completion, the ELT will embark on its mission to:

- Directly image and analyze the atmospheres of Earth-like exoplanets , searching for signs of habitability and life.
- Witness the Cosmic Dawn, observing the formation of the universe's first stars and galaxies.
- Probe the mysteries of dark matter and dark energy by studying the universe's expansion and the growth of cosmic structures. Watch stars being born and planets coalescing from disks of dust and gas.

Figure 4: Comparison of images of crowded stellar fields observed by the Hubble Space Telescope (HST, left), the James Webb Space Telescope (JWST, centre), and ELT's MICADO instrument (right) for three different stellar densities. Credit:

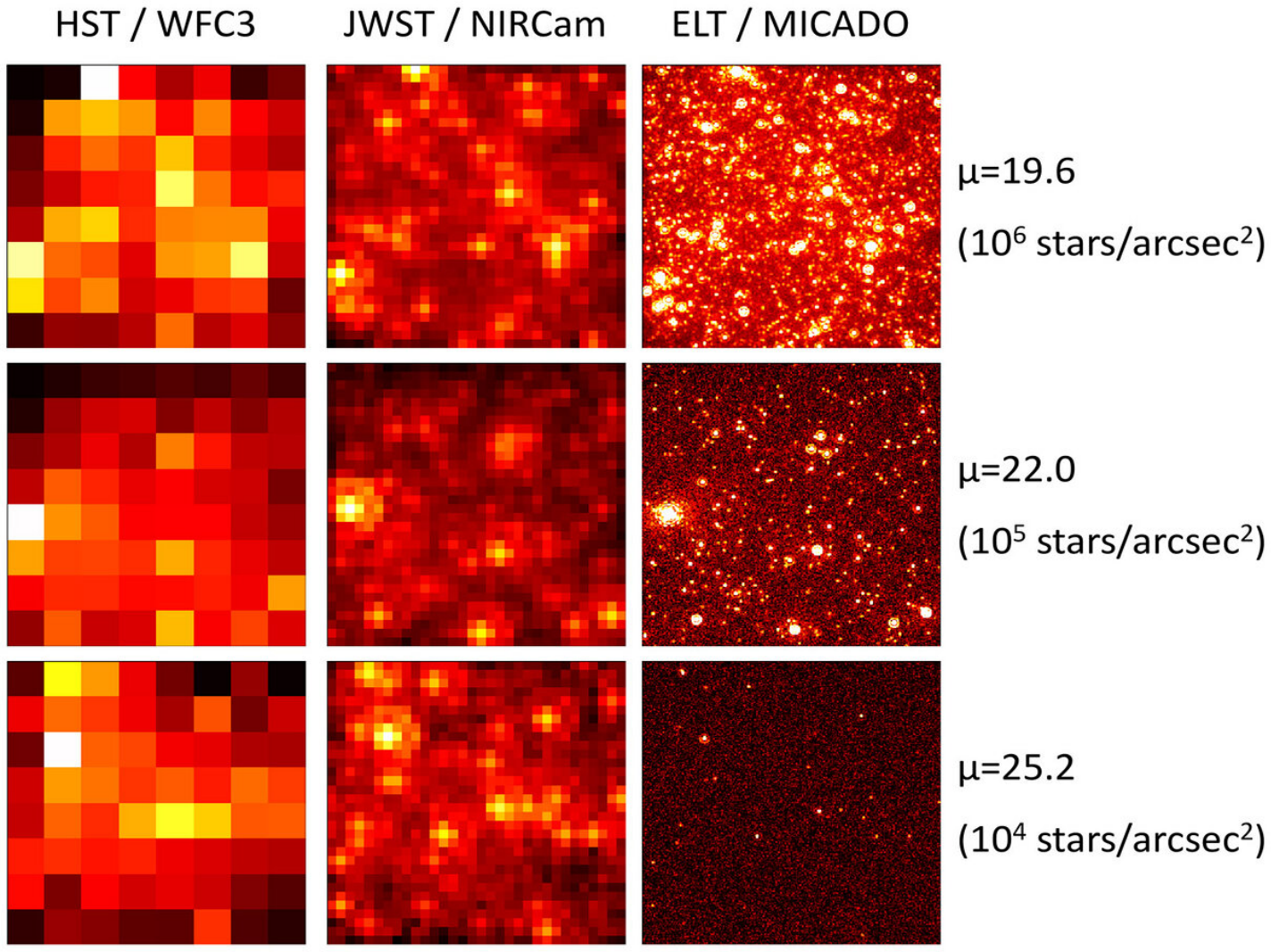

Among the ELTs, the ELT is the largest, GMT is the light-collecting specialist, but TMT is the highest-precision imaging and astrometry machine. The Extremely Large Telescope is more than an instrument; it is a gateway. It represents humanity's relentless curiosity, poised to open a new window on the universe and answer questions we have only just begun to ask.

## 3 Conclusion

The ELT has currently opted for a standard protected silver coating, which challenges its near-UV performance. The ELT will not work alone. It will be part of an unparalleled era of discovery, teaming up with observatories like the GMT, TMT, James Webb Space Telescope, the future Roman Space Telescope and many more telescopes. A sequel to this article will describe the GMT and TMT.

# 4 Acknowledgements